# HARDWARE ACCELERATION OF THE GIPPS MODEL FOR REAL-TIME TRAFFIC SIMULATION

Salim Farah[1] and Magdy Bayoumi[2]

The Center for Advanced Computer Studies, University of Louisiana at Lafayette, USA
[1] snf3346@cacs.louisiana.edu
[2] mab@cacs.louisiana.edu

## ABSTRACT

*Traffic simulation software is becoming increasingly popular as more cities worldwide use it to better manage their crowded traffic networks. An important requirement for such software is the ability to produce accurate results in real time, requiring great computation resources. This work proposes an ASIC-based hardware accelerated approach for the AIMSUN traffic simulator, taking advantage of repetitive tasks in the algorithm. Different system configurations using this accelerator are also discussed. Compared with the traditional software simulator, it has been found to improve the performance by as much as 9x when using a single processing element approach, or more depending on the chosen hardware configuration.*

## 1. INTRODUCTION

The steady improvement in computation power has allowed for many applications previously limited only to super-computers and data centers. Traffic simulation is one application that has begun to gain popularity in recent years, especially in cities with notoriously busy traffic networks such as Madrid and Singapore. The technology is used in either the design or operation phase of a transportation network. During design, simulation helps decide upon the most efficient and reliable configuration, while simulation during traffic operation allows for predicting rush hours and traffic flow, as well as effective rerouting in case of road closures.

### 1.1 Traffic Simulators Overview

Traffic simulators are handed the system data from road traffic sensors, and already have the information about the road network and its layout. Simulating traffic flow can be performed in three methods. In the macroscopic method, the traffic system is modeled at no lower than the road level and the density at the given road stretches. From there the flow development is carried out [1]. On the other hand, the microscopic method simulates at the car level, following each car's movement. Combining all the individual car behaviors, the overall traffic flow can be obtained. In the middle sits the mesoscopic method which is a trade-off between the two. The microscopic method offers the greatest accuracy at the expense of computation time, while the macroscopic approach will produce a less accurate result in a more timely fashion.

A number of traffic simulators exist, some of which are commercially sold and widely used in a variety of fields. These simulators have been evolving for a while, and it's safe to say they've reached a mature state where they can perform at a reasonable speed and produce trusted results. AIMSUN and VISSIM [2] are two popular commercial traffic simulators used by a number of traffic engineering firms and transportation planning agencies. AIMSUN is allegedly capable of microscopic simulation of traffic in a big-sized city with a speed 60x faster than that of real time, or in other words simulates 1 hour in 1 minute.

## 1.2 Related Work

The previously discussed simulators are purely software-based, as are most implementations. However, there have been a few proposed FPGA-based implementations. One work combines microprocessors with FPGAs in a low-bandwidth, high-latency interconnect, dividing the tasks between software and hardware to balance the workloads [3]. It claims to reduce the number of needed FPGAs and to achieve a speedup of 12.8x over an AMD processor. Another FPGA design centers around modeling the route system as a collection of interconnected cells, each cell representing a short segment of a roadway and can be either empty or containing a single car [4]. The authors claim to be able to effectively model various geometric configurations of a traffic system by hierarchically combining the cells.

## 1.3 Performance Requirements

A timely response is the obvious prime requirement of real-time simulation systems. Delays cannot be tolerated in a traffic environment. Once a traffic congestion has been formed, it is hard to reverse it due to the unidirectional nature of vehicle movement, and on busy highways congestions can form in a matter of seconds in a major incident. Reacting as quickly as possible is therefore a must, and this entails a very efficient simulation of any decision the system might decide to take.

An important thing to note is that upon deciding on a response strategy, a potentially large number of possible actions are simulated. Having a computing infrastructure capable of parallel processing is therefore desirable. Still, some of the simulations may be related to each other and depend on each other's results, and parallelizing them may not be possible. Simulation runs will in this case add up in time and a 1 minute run would add up to 15 minutes if performed in 15 different instances.

## 2. PROPOSED HARDWARE ACCELERATION

### 2.1. Choice of Simulation Model

Traffic simulators usually make use of two important models: the car-following model and the lane-switching model. This work is only concerned with the car-following model, although it could be extended to the lane-switching model if desired. The car-following model itself is modeled differently in different simulators. VISSIM and AIMSUN are two of the most widely used traffic simulators, and they employ different car-following models. The AIMSUN model was ultimately chosen for the hardware acceleration, for the two following reasons.

Firstly, the AIMSUN simulator uses the Gipps car-following model, a model represented mathematically through an algebraic equation, as opposed to the model used in VISSIM, which is based on a psychological model that tries to mimic the behavior of the driver [5]. Mapping a mathematical equation to hardware is more straightforward than trying to accelerate a complex psychological model that relies on statistical decisions and specialized algorithms. The mathematical nature of the Gipps model in AIMSUN allows the use of common hardware units such as dividers and multipliers, and the reuse of such items in case the hardware unit is to be used to accelerate other tasks as well.

The second reason for choosing AIMSUN is simply its superior accuracy, as was concluded in [5]. Accuracy is not to be underestimated in traffic simulators as errors would accumulate considering cars are all affected by each others' movements. AIMSUN was found to produce fewer errors when taking real life situations as a reference.

The following is the primary equation representing the Gipps car-following model:

$$V_a(n, t + T) = V(n, T) + 2.5\alpha(n)T \times (1 - \frac{V(n,T)}{V^*(n)})\sqrt{(0.025 - \frac{V(n,T)}{V^*(n)})}, \quad (1)$$

where
$V(n, t)$ is the speed of vehicle n at time t,
$V*(n)$ is the desired speed of vehicle n,
$a(n)$ is the maximum acceleration for vehicle n,
$T$ is the reaction time (this is equal to simulation step).

The actual model is a little more complicated than this, it uses an additional equation for calculating the velocity value and then chooses the lower of the two values from the two equations. However only the above equation will be accelerated here, but the same concept can be applied to that second equation. This work hopes to show the potential of accelerating the Gipps model, more so than performing a full-fledged acceleration. This is the reason why only part of the model was accelerated.

**2.2 Assumed System Organization**

It was assumed that only the equation above was moved to hardware while the rest of the software constituting the simulator is still unchanged. That is the code in the simulator responsible for carrying out the calculation above is now replaced by one instruction that performs this task. This instruction requires 4 operands: acceleration, time step, desired velocity, and current velocity. These can be stored in special memory registers or locations prior to executing the instruction.

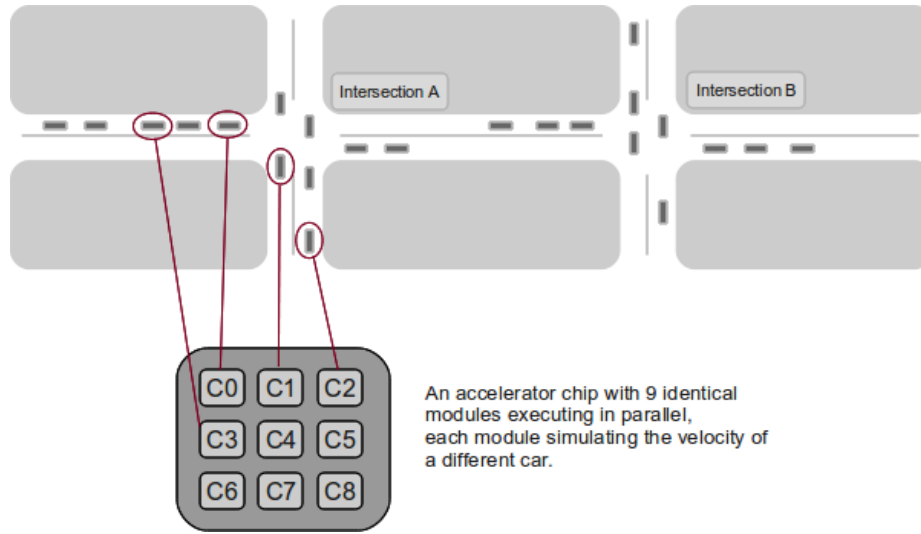

Figure 1. Illustration of how an array of accelerators can concurrently simulate different cars

The hardware accelerator module can be either added to the internals of the general purpose CPU, or installed as an add-on via some fast connection protocol like PCI Express. In the latter solution the module would occupy the whole chip, therefore giving it a much larger area and power budget. This in turn allows for the use of multiple modules on a single chip, performing parallel calculations that correspond to modeling several cars simultaneously. The simulation would thus be accelerated by several orders of magnitude, even when accounting for the communication overhead due the bus CPU connection.

## 2.3 Hardware Architecture

As is evident in the equation discussed above, the hardware units that will be needed are an adder, a multiplier, and a divider. Given these units, it's still required to perform the square root operation. The multiplier and divider units shall be discussed first, and subsequently the square root implementation shall be dealt with. But before that, it's worth discussing the word width that is being assumed. Given that the maximum speed anyone is likely to achieve is below 256 km/h, 8 bits should be assigned to represent both the speed and the acceleration. However, given that the equation also deals with decimal points, 6 more bits were assigned as fixed decimal point bits. These allow for an accuracy of 0.0156, i.e. an error of ± 0.008, which should be acceptable for the given application. In all, 14 bits were used to represent the numbers in use.

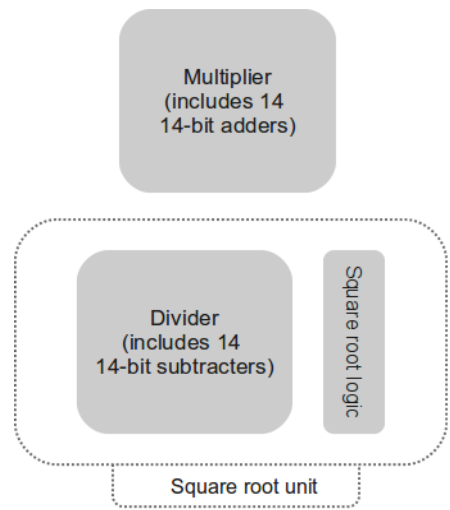

Figure 2. The high level organization of the hardware accelerator

For the multiplier, an entirely combinational approach was chosen, since the main target of this work is speed and performance. With a 14-bit word, the multiplier had to use 14 14-bit adders, adding up to 196 full adder cells. The same goes for the divider, which uses 14 14-bit subtracters, which are essentially adders with an added inverter on one of the inputs. Obviously, significant area is being occupied by just the multiplier and the divider. But these two units take the vast majority of the design and everything else takes insignificant area in comparison. For instance the square root unit already makes use of the available divider, adding only little hardware to that, and the control unit for the design is also small in comparison. Moreover, it should be kept in mind that we're mainly assuming the accelerator will be on a chip by itself.

The Babylonian method was used for the computation of the square root. Initially a rough estimation based on the number of bits to the right of the first '1' in the number is done, giving a starting point close to the solution so that the unit would converge much faster. Subsequently, the following operation is conducted multiple times until it converges to a constant number:

$$x_{n+1} = \frac{1}{2}(x_n + \frac{S}{x_n}), \qquad (2)$$

where S is the number the square root of which is being sought. It was observed that it only took two clock cycles (or iterations) at most for the operation to converge to a constant value, due mainly to a relatively accurate estimation done beforehand.

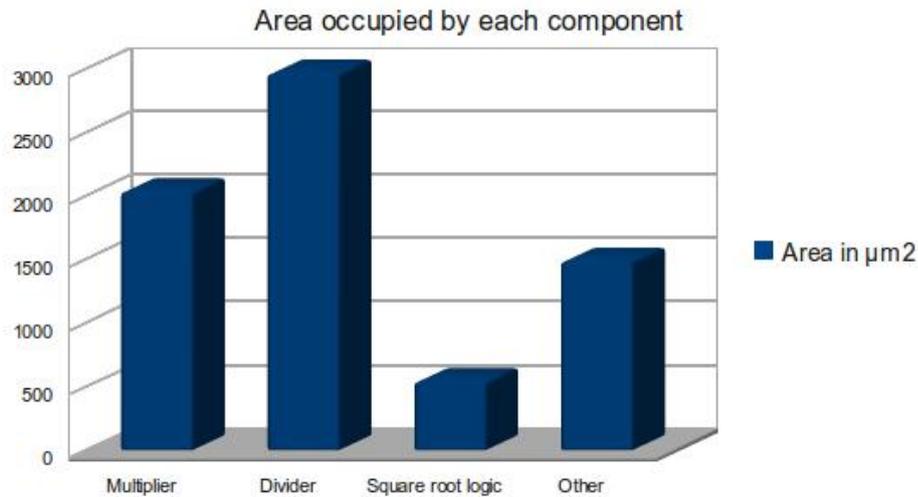

Figure 3. Area distribution of the accelerator. Total area was 7016 μm2.

## 3. RESULTS

The Verilog code for the accelerator was synthesized with a 45 nm standard cell library (FreePDK45). The operating frequency was set to be 250 MHz, which is about the highest the design could reach without timing violations. Although this may seem like too low for such an advanced technology, the entire computation only takes 4 clock cycles, or 16 ns. If desired, however, the clock frequency may be significantly increased if instead of entirely combinational dividers and multipliers, hybrid ones were used that take a few clock cycles to complete one division or multiplication operation. That would also dercease the area of these units due to the reuse of the adders/subtracters. When considering packing several of these accelerators on one chip, this becomes especially important. The implemented design occupies an area of 7016 μm2 with an estimated power consumption of 2.3 mW.

This result was compared with an estimation of the time required to complete the computation of the Gipps model equation on a general purpose processor. A short program that performs the same computation was written in C, and was run on an Intel Core i3-350M processor, which is a mid-range dual core processor with 3 MB of cache, and 2 threads per core, making a total of 4 virtual cores. The computer was running a Linux 64-bit OS, and has a total of 4 GB of RAM. Code profiling functions were added to the C program to measure the execution time, and the computation was run for 100 iterations in order to average out any inaccuracies in the profiling measurement. The average execution time was 144 ns, which is 9x slower than the hardware accelerator. When using multiple processing elements of the accelerator, this speedup would be multiplied by the number of PEs in use.

## 4. CONCLUSION

The use of hardware accelerators for improving the performance of the AIMSUN traffic simulator has been shown to be significantly effective. The hardware accelerator uses high performance multiplication and division units, and is able to perform an accurate square root operation in only two clock cycles. By comparison, the software code written in C and performing the same computation was 9x slower. The obtained speedup would in fact be multiplied when the accelerator includes several computation units working in parallel, which is feasible when the accelerator is implemented off-chip. Future work could concentrate on finding an efficient way for using and placing an array of acceleration modules working concurrently on a single chip.